\documentclass[12pt,preprint,showpacs,preprintnumbers,amsmath,amssymb]{revtex4}

\usepackage{graphicx}
\usepackage{dcolumn}
\usepackage{bm}

\begin{document}

\title{Stable Control of Pulse Speed in Parametric Three-Wave Solitons}

%
%
%
%
%
%
%
%
%

\author{Antonio Degasperis,$^1$ Matteo Conforti,$^2$ Fabio Baronio,$^2$ and Stefan Wabnitz$^3$ }
\affiliation{$^1$Dipartimento di Fisica, Istituto Nazionale di
Fisica Nucleare, Universit\`a ``La Sapienza'', 00185 Roma, Italy\\
$^2$Dipartimento di Elettronica per l'Automazione, Universit\`a di
Brescia,  25123 Brescia, Italy\\
$^3$Laboratoire de Physique, Universit\'e de Bourgogne, UMR CNRS
5027,  21078 Dijon, France}

\date{\today}

\begin{abstract}
We analyze the control of the propagation speed of three wave
packets interacting in a medium with quadratic nonlinearity and
dispersion. We found analytical expressions for mutually trapped
pulses with a common velocity in the form of a three-parameter
family of solutions of the three-wave resonant interaction. The
stability of these novel parametric solitons is simply related to
the value of their common group velocity.

\end{abstract}

\pacs{05.45.Yv, 42.65.-k, 42.65.Sf, 42.65.Tg, 52.35.Mw}
\maketitle

A three-wave resonant interaction (TWRI) is ubiquitous in various
branches of science, as it describes the mixing of waves with
different frequencies in weakly nonlinear and dispersive media.
Indeed, TWRI occurs whenever the nonlinear waves can be considered
as a first-order perturbation to the linear solutions of the
propagation equation. TWRI has been extensively studied alongside
with the development of nonlinear optics, since it applies to
parametric amplification, frequency conversion, stimulated
Raman and Brillouin scattering. In the context of plasma physics,
TWRI describes laser-plasma interactions, radio frequency heating,
and plasma instabilities. Other important domains of application
of TWRI are light-acoustic interactions, interactions of water
waves, and wave-wave scattering in solid state physics. Two
classes of analytical soliton solutions of the TWRI have been
known for over three decades. The first type of solitons describes
the mixing of three pulses which travel with their respective
linear group velocity, and interact for just a short time
\cite{armstrong70,zakharov73,kaup76,degasperis06}. The second type
of solitons, also known as simultons, are formed as a result of
the mutual trapping of pulse envelopes at the three different
frequencies. Hence the three wave packets travel locked together
with a common group velocity \cite{nozaki74}. In all of the above
discussed domains of application, parametric TWRI solitons play a
pivotal role because of their particle-like behaviour, which
enables the coherent energy transport and processing of short wave
packets \cite{armstrong70,nozaki74,ibragimov96}.

In this Letter we reveal that the class of TWRI simultons (TWRIS)
is far wider than previously known. We found a whole new  family
of bright-bright-dark triplets that travel with a common, locked
velocity. The most remarkable physical property of the present
solitons is that their speed can be continuously varied by means
of adjusting the energy of the two bright pulses. We studied the
propagation stability of TWRIS and found that a stable triplet
loses its stability as soon as its velocity decreases below a well
defined critical value. Another striking feature of a TWRIS is
that an unstable triplet decays into a stable one through the
emission of a pulse, followed by
acceleration  up to reaching a stable velocity. 

The coupled partial differential equations (PDEs) representing TWRI in (1 + 1)
dimensions read as \cite{zakharov73}:
\begin{eqnarray}\label{3wri}
\nonumber E_{1t}-V_1 E_{1z}&=&\phantom{-} E_2^*E_3^*,\\ E_{2t}-V_2
E_{2z}&=&-E_1^*\,E_3^*,\\ \nonumber E_{3t}-V_3
E_{3z}&=&\phantom{-} E_1^*\,E_2^*,
\end{eqnarray}
where the subscripts $t$ and $z$ denote derivatives in  the
longitudinal and transverse dimension, respectively. Moreover,
$E_n=E_n(z,t)$ are the complex amplitudes of the three waves,
$V_n$ are their linear velocities, and $n=1,2,3$. We assume here
the ordering $V_1>V_2>V_3$ which, together with the above choice
of signs before the quadratic terms, entails the non--explosive
character of the interaction. In the following, with no loss of
generality, we shall write the equations (\ref{3wri}) in a
reference frame such that $V_3=0$. A remarkable property of the
Eqs.(\ref{3wri}) is their invariance with respect to the
transformation
\begin{equation}\label{inv}
\hat E_n(z,t)=s\exp[i(q_nz_n+\alpha_n)] E_n(sz + z_0,st+t_0)
\end{equation}
where $\alpha_1+\alpha_2+\alpha_3=0$, $q_n=q(V_{n+1}-V_{n+2})$,
$z_n=z+V_n t$ are the characteristic coordinates and $n=1,2,3 \,
mod(3)$. As the transformation (\ref{inv}) depends on six real parameters, namely
$\alpha_1, \alpha_2, s, q, z_0$ and $t_0$, clearly one may
introduce these parameters in the expression of any given
solution of the TWRI equation.

The evolution equations (\ref{3wri}) represent an infinite-dimensional Hamiltonian
dynamical system, with the conserved Hamiltonian
\begin{equation}\label{H}
\begin{split}
H &=\frac{1}{4i}\int_{-\infty}^{+\infty} \bigg[
     V_1(E_{1z}^*E_1-E_{1z}E_1^*)
     - V_2(E_{2z}^*E_2-E_{2z}E_2^*)\\
+&V_3(E_{3z}^*E_3-E_{3z}E_3^*)
  + 2E_1E_2E_3 - 2E_1^*E_2^*E_3^*
 \bigg]dz,
\end{split}
 \end{equation}
energies (Manley-Rowe invariants)
\begin{equation}\label{E12}
 I_{12}=I_1+I_2=\frac{1}{2}\int_{-\infty}^{+\infty} ( |E_1|^2 + |E_2|^2) dz,
 \end{equation}
\begin{equation}\label{E23}
 I_{23}=I_2+I_3=\frac{1}{2}\int_{-\infty}^{+\infty}( |E_2|^2 + |E_3|^2 ) dz,
\end{equation}
and total transverse momentum
\begin{equation}\label{J}
\begin{split}
J=\frac{1}{4i}\int_{-\infty}^{+\infty}[(&E_1^*E_{1z}-E_1E_{1z}^*
)-(E_2^*E_{2z}-E_2E_{2z}^* )\\ +(&E_3^*E_{3z}-E_3E_{3z}^* ) ]dz.
\end{split}
\end{equation}
Each of the above conserved quantities is related to a given
internal parameter of the TWRIS which, in turn, is associated with
a symmetry (e.g., phase rotation or space translation) of the TWRI
equations (\ref{3wri}) \cite{buryak02}. As a consequence, one may
expect that Eqs.(\ref{3wri}) possess a three-parameter family of
soliton solutions. We found such soliton solutions by using
the recent results on the TWRI equations as presented in
Ref.\cite{calogero05}. Their expression is
\begin{subequations}\label{simult-}
\begin{equation}\label{E1}
E_1=\frac{2p \, a^*}{\sqrt{|b|^2+|a|^2}} \frac{g_1}{g(V_1-V_2)}
\frac{\exp[i(q_1z_1-\chi z+\omega t )]}{\cosh[B(z+Vt )]},
\end{equation}
\begin{equation} \label{E2}
E_2=\frac{-2p \, b}{\sqrt{|b|^2+|a|^2}}\frac{g_2}{g(V_1-V_2)}
\frac{\exp[i(q_2z_2+ \chi  z-\omega t)]}{\cosh[B(z+Vt )]},
\end{equation}
\begin{equation} \label{E3}
E_3=\{1+\frac{2p \, b^*}{|b|^2+|a|^2} [1-\tanh[B(z+Vt )]]\} \,
\frac{a \, g_3 \exp(iq_3z_3)}{g(V_1-V_2)}\\
\end{equation}
\end{subequations}
where
\begin{gather*}\label{def}
b  = (Q -1)(p + i k/Q),\ \ \ \ \ r=p^2-k^2-|a|^2, \\
 Q= \frac{1}{p}\sqrt{ \frac12 [\,\,r + \sqrt{r^2
 +4k^2p^2}\,\,]},\\
 B=  p[\,V_1+V_2 - Q (V_1-V_2)\,]/(V_1-V_2),\\
 V=  2V_1 V_2/ [\,V_1+V_2 - Q (V_1-V_2)\,],\\
 \chi = k [\,V_1+V_2 -(V_1-V_2)/Q\,]/(V_1-V_2),\\
 \omega =  -2k V_1 V_2/(V_1- V_2),\ \ \ q_n=q(V_{n+1}-V_{n+2}),\\
g_n=|(V_{n}-V_{n+1})\,(V_{n}-V_{n+2})|^{-1/2}\,\,,\,\,g=g_1\,g_2\,g_3\,\,,\\
%
%
\end{gather*}
and $n=1,2,3 \ mod \, (3)$. The above
expressions depend on the five real parameters $V_1,V_2,p,k,q$, and
the complex parameter $a$.
From the definition of $Q$, one can see that the
above parameters must be chosen so that if $k=0$, then $p^2>|a|^2$.

The TWRIS is composed of two bright pulses (\ref{E1}, \ref{E2}),
and a kink or shock-like pulse (\ref{E3}), which travel with a
common locked velocity $V$. The expressions (\ref{simult-}) may be
represented in a more convenient form as
\begin{equation}\label{sim2}
E_n(\xi,\tau)=U_n(\xi)\exp[\,i\Phi_n(\xi,\tau)\,], \,\, n=1,2,3.
\end{equation}
Here we used a reference frame that moves along with the soliton,
with co--ordinates $\xi=z+Vt,\tau=t$ where $U$ and $\Phi$ are real
functions and $\Phi_n(\xi,\tau)=\phi_n\tau+f_n(\xi)$.
\begin{figure}
 \begin{center}
     \includegraphics[width=6cm]{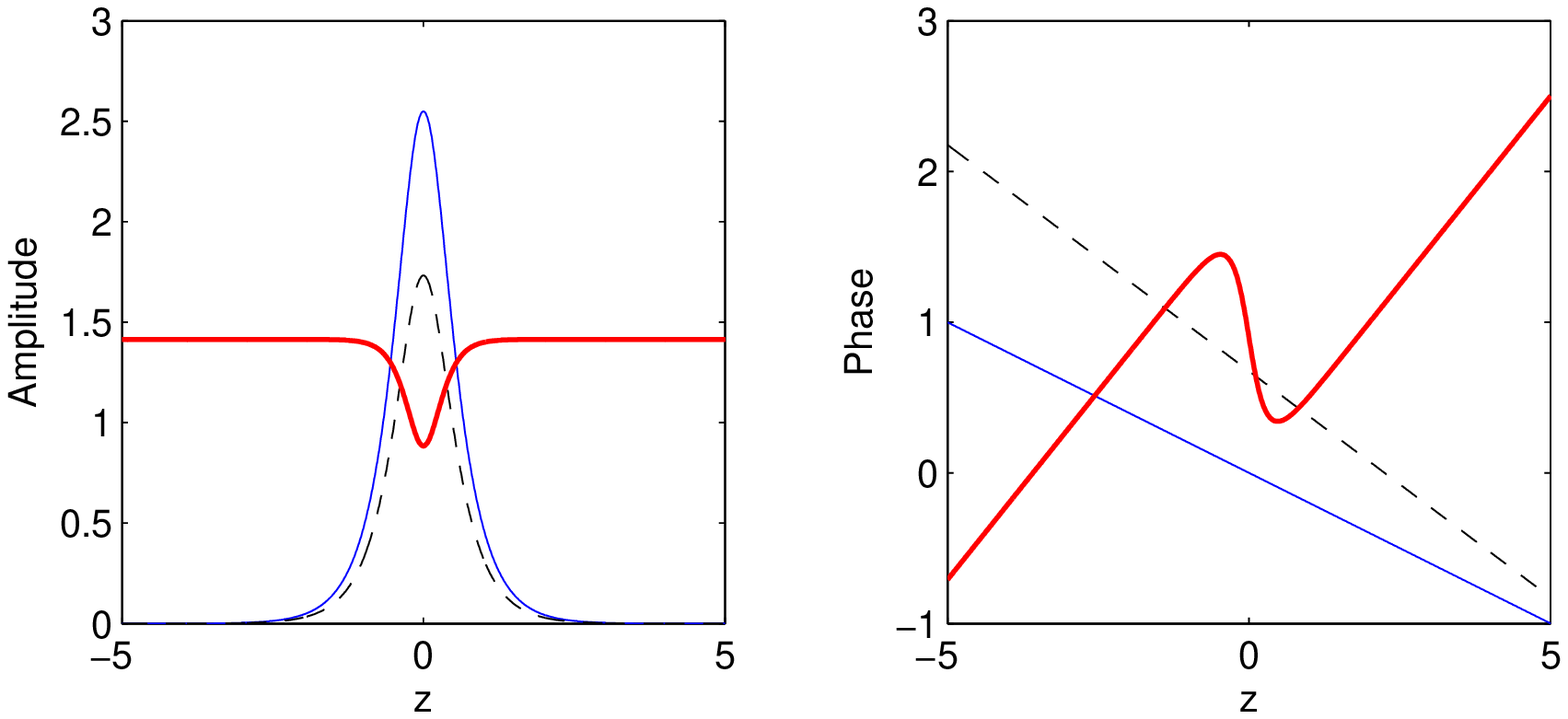}\\
     \includegraphics[width=6cm]{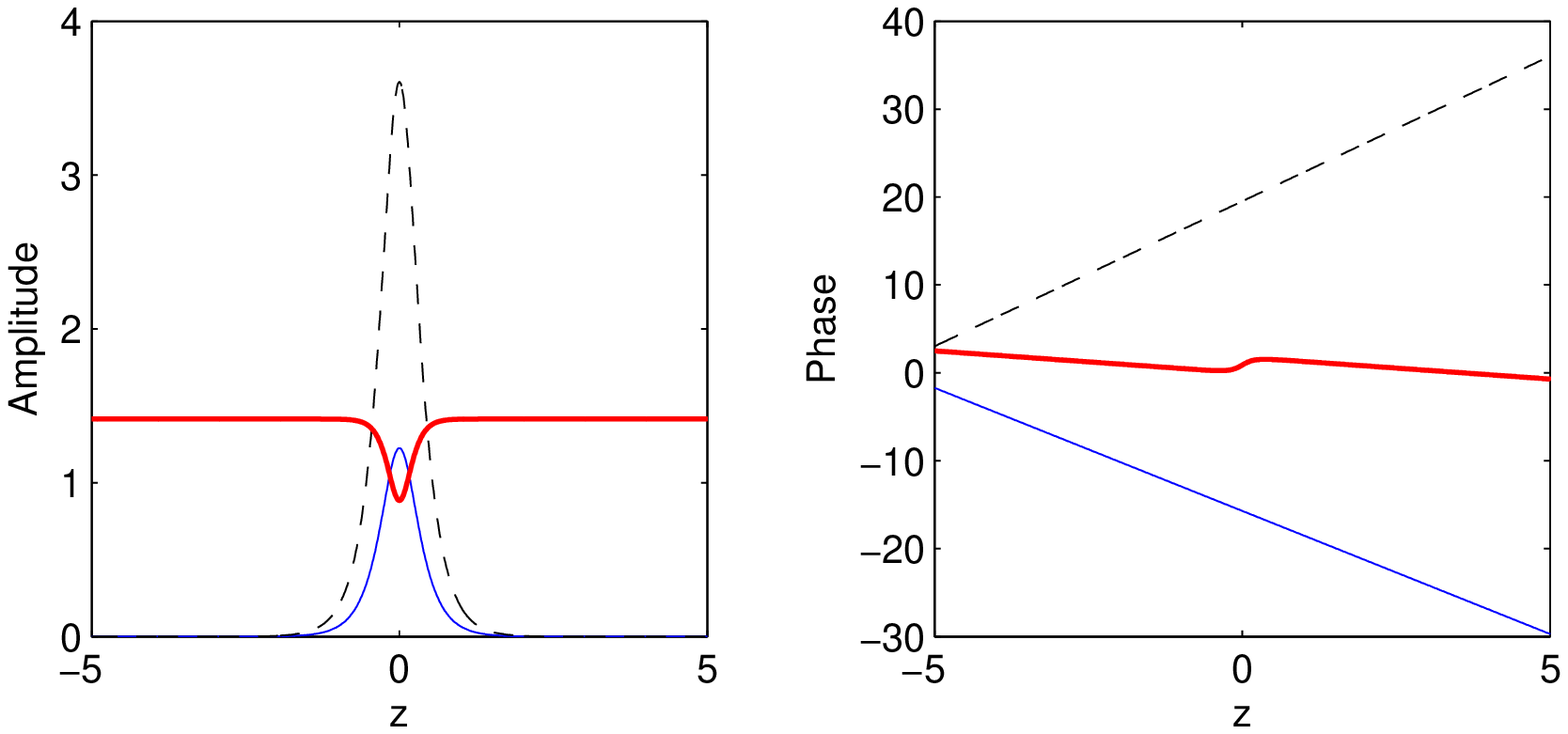}
        \end{center}
    \caption{(color online). Amplitude and phase of two simultons (\ref{simult-}) at $t=0$ with $V_1=2$, $V_2=1$, $a=1$, $k=0.5$, $q=0.5$; in the upper figures $p=1$, in the lower figures $p=-1$.
    Thin solid curve $E_1$, dashed curve $E_2$, thick solid curve $E_3$.}\label{profili}
\end{figure}
A simple analysis of (\ref{sim2}) shows that, for any value of the
parameters, the pulse amplitudes $U_1(\xi)$, $U_2(\xi)$ and
$U_3(\xi)$ are even functions of $\xi$ and the phase constants
satisfy $\phi_1+\phi_2+\phi_3=0$. On the other hand, if $k=0$ the
phase profiles are all piecewise linear in $\xi$ and obey the
condition $f_1(\xi)+f_2(\xi)+f_3(\xi)=0$ for $p \xi >0$ and
$f_1(\xi)+f_2(\xi)+f_3(\xi)=\pi$ for $p \xi <0$. Whereas for $k \neq
0$ the phase profile $f_3(\xi)$ is nonlinear and
$\cos[f_1(\xi)+f_2(\xi)+f_3(\xi)]$ is an odd function of $\xi$;
moreover the kink pulse $E_3$ is ``grey'' if $k\neq 0$ and is
``dark'' if $k=0$. Such amplitude and phase front profiles prevent
a net energy exchange among the three waves. It is important to
point out that the condition $-1<Q<1$ leads to a speed $V$ that
lies in-between the characteristic velocities $V_1$ and $V_2$ of
the two bright pulses, i.e. $V_1>V>V_2$. The above described
properties mean that TWRIS represent a significant generalization
with respect to previously known three-wave solitons which exhibit
a simple (constant) phase profile and correspond to the special
case $k=q=0\,,\,r>0$ \cite{nozaki74}. In
Fig.\ref{profili} we plotted two characteristic examples of TWRIS
amplitude and phase-fronts (\ref{simult-}).

It is interesting to consider the physical meaning of the various
TWRIS parameters appearing in (\ref{simult-}). For a given choice
of the characteristic linear velocities $V_1$ and $V_2$, we are
left with the four independent parameters $p,k,q$, and $a$. We may
note that $p$ is basically associated with the scaling of the wave
amplitudes, as well as of the coordinates $z$ and $t$. The
parameter $a$ determines the amplitude of the asymptotic plateau
of the kink $E_3$. The value of $k$ provides the wave--number of a
``carrier--wave''. The parameter $q$ adds a phase contribution
which is linear in $z$ and $t$. Since the system (\ref{3wri}) is
invariant under a transformation (\ref{inv}), without loss of
generality we may set $a=1$, which reduces the number of essential
parameters to just three, corresponding to the three symmetries of
Eqs.(\ref{3wri}). The parameters $p,k,q$ in (\ref{simult-}) may be
more conveniently  mapped into the parameters $V,\phi_1,\phi_2$ of
Eq.(\ref{sim2}), which provide a more direct physical insight
into the features of a TWRIS. Such a mapping is obtained by
comparing Eqs.(\ref{simult-}) with (\ref{sim2}), and reads as: $ V=
2V_1 V_2/ [\,V_1+V_2 - Q (V_1-V_2)\,], \,\, \phi_1=
qV_2(V_1-V)+\chi V+\omega, \,\, \phi_2= qV_1(V-V_2)-\chi V-\omega
$. The TWRIS is thus simply expressed in terms of its velocity $V$
and the two phase constants $\phi_1$ and $\phi_2$.
\begin{figure}
 \begin{center}
     \includegraphics[width=3.5cm]{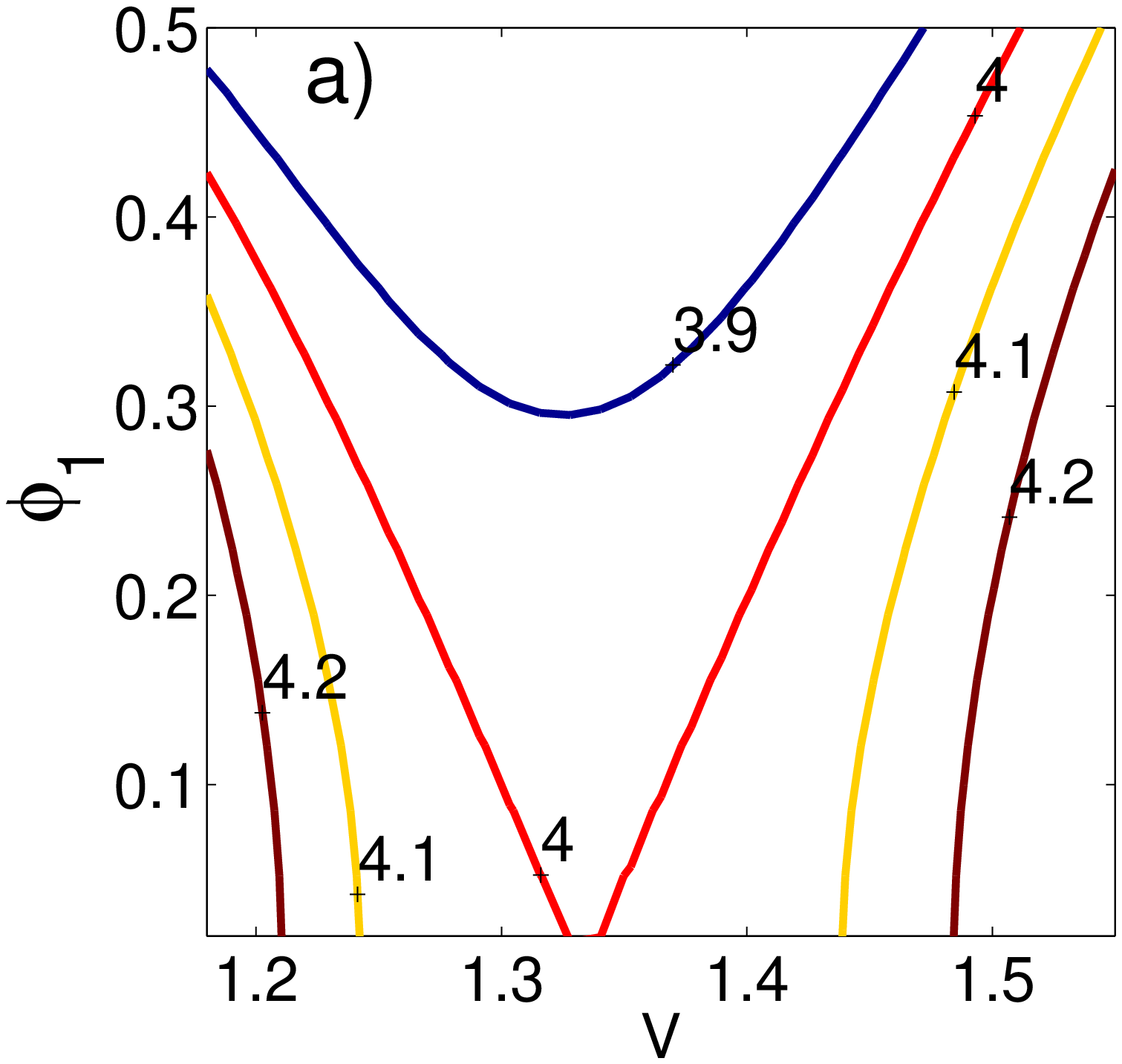}\,\,\,\,\,\,\,\,\,\,\,
     \includegraphics[width=3.9cm]{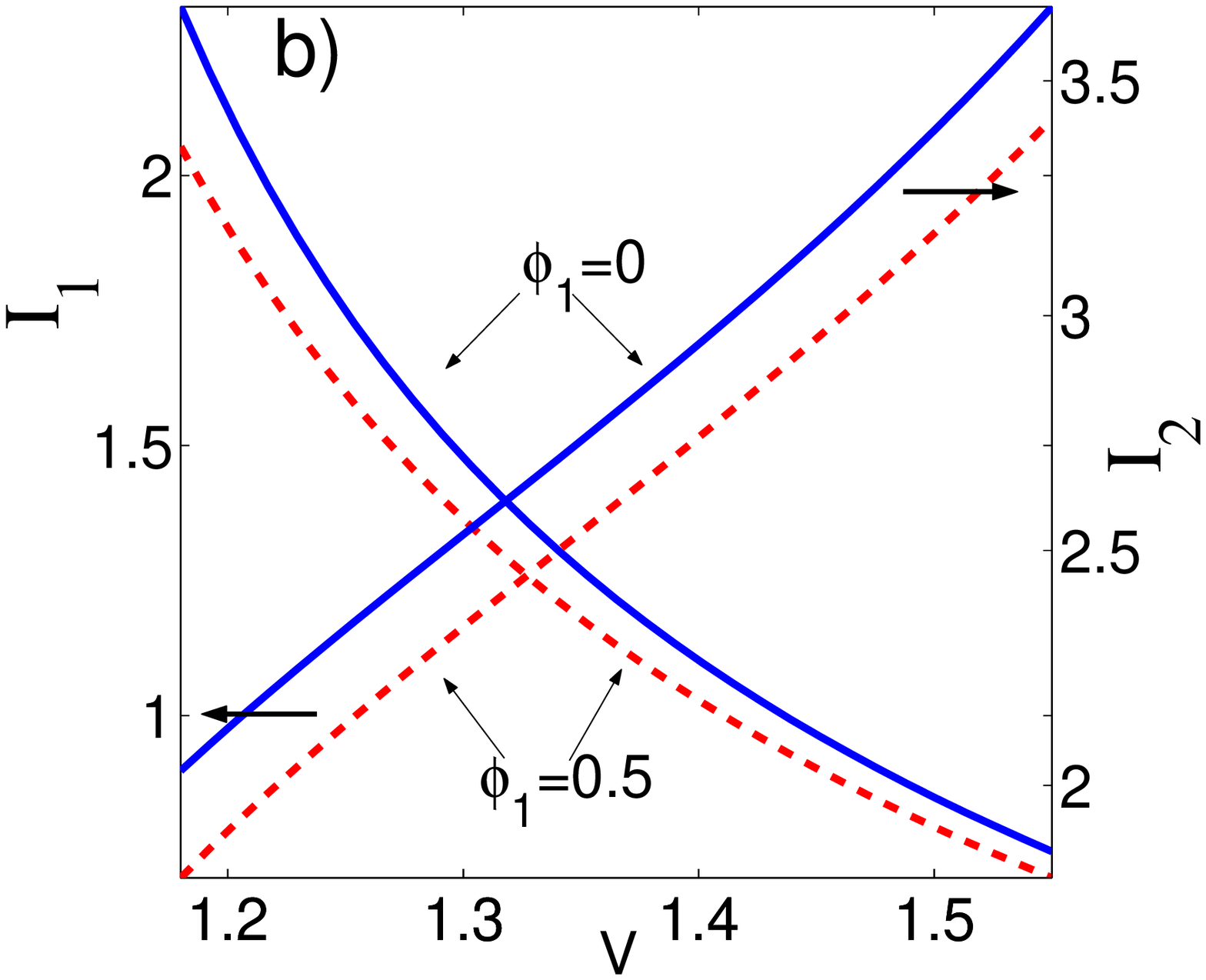}
        \end{center}
    \caption{(color online). (a)  phase constant $\phi_1$ versus velocity $V$
    for different energies $I_{12}.$ (b) energies $I_{1}$, $I_2$
    versus velocity $V$ for representative values of the phase constant $\phi_1$. In all
cases $\phi_2=-\phi_1$, $V_1=2$, $V_2=1$.}\label{andamenti}
\end{figure}
Let us investigate what are the TWRIS properties for a fixed
choice of the linear velocities $V_1$ and $V_2$, upon variations
of its energy flows and transverse momentum.  As an example, Fig.
\ref{andamenti}(a) shows the dependence of the phase constant
$\phi_1$ on the locked velocity $V$, for the case where
$\phi_2=-\phi_1$, with different values of the conserved energy
$I_{12}$. Moreover, Fig. \ref{andamenti}(b) illustrates the
dependence of the energies $I_1$ and $I_2$ (which happen to be
time--independent for a TWRIS) on the locked velocity $V$, for
different choices of the phase constant $\phi_1=-\phi_2$. As it
can be seen, the intensity and phase profiles, as well as the
energy distribution among the different wave packets, strongly
depend upon the value of the locked velocity $V$.

\begin{figure}
 \begin{center}
     \includegraphics[width=8cm]{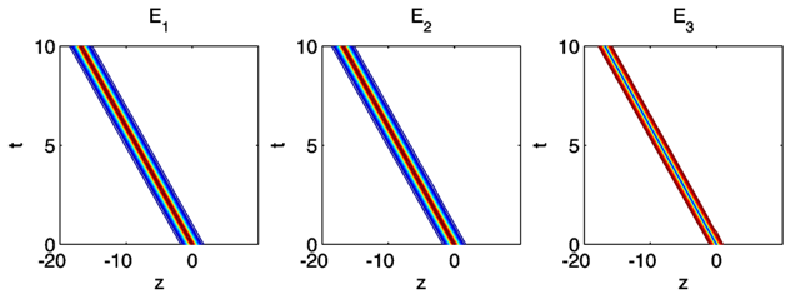}\\
     \includegraphics[width=8cm]{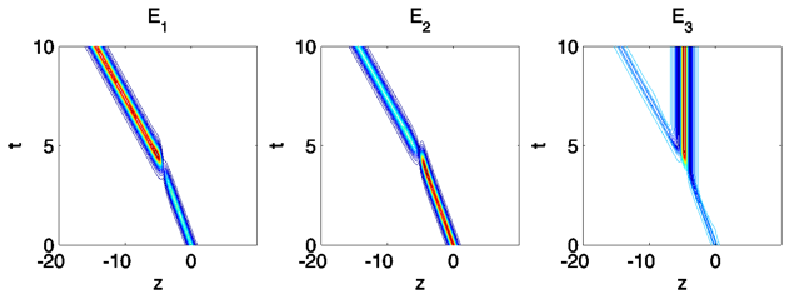}
 \end{center}
    \caption{(color online). Propagation of stable and unstable simultons.
    The common parameters are $V_1=2$, $V_2=1$, $a=1$, $k=0.5$, $q=1$. For the stable simulton (upper figures) $p=1$
($V=1.68>V_{cr}$, $\phi_1=0.5$, $\phi_2=2.2$),
    whereas for the unstable simulton (lower figures) $p=-1$ ($V=1.1<V_{cr}$, $\phi_1=1.43$, $\phi_2=-0.33$).}\label{runs}
\end{figure}

The next crucial issue is the propagation stability of
TWRIS. A first insight into this problem may be provided by
performing a  linear stability analysis (LSA) as in
Ref.\cite{conforti05}. Let us consider a perturbed TWRIS of the
form
 $$ \tilde
E_n(\xi,\tau)=(x_n(\xi)+P_n(\xi,\tau))e^{i\phi_n\tau}, \quad
n=1,2,3$$ where $x_n(\xi)=U_n(\xi)\exp[{if_n(\xi)}]$ is the
soliton profile, and we consider a weak perturbation
$|P_n|<<|x_n|$. By inserting the above ansatz in Eqs.(\ref{3wri}),
and by retaining only linear terms in $P_n$, one obtains a linear
system of PDEs. For the numerical analysis, these PDEs can be
reduced to a system of ordinary differential equations $\dot
P(\tau)=MP(\tau)$, by approximating the spatial derivatives with
finite differences, where $P$ is the perturbation vector sampled
on a finite grid. A necessary condition for the instability of a
stationary solution $x_n(\xi)$ is that the matrix $M$ has at least
one eigenvalue with positive real part. Numerical computations
over a wide parameter range show that eigenvalues of $M$ exist
with a positive real part whenever $p<0$. On the other hand, for
$p>0$ the largest real part of the eigenvalues is equal to zero,
which means that the TWRIS are only neutrally stable.
Note that the instability
condition $p<0$ leads to the inequality
$V<V_{cr}=2V_1V_2/(V_1+V_2)$.
Extensive numerical integrations of Eqs.(\ref{3wri}) confirm that
TWRIS with $V<V_{cr}$ ($V>V_{cr}$) are always unstable (stable).
The propagation of  either stable or unstable  TWRIS is
illustrated in Fig. \ref{runs}, which shows the general feature of
unstable solitons with $V<V_{cr}$. Namely, the simulton decays
into a stable soliton with $V>V_{cr}$, and it emits a pulse in the
wave $E_3$. It is quite remarkable that the dynamics of the decay
from unstable into stable solitons may be exactly described by
analytical solutions with variable velocity or boomerons
\cite{calogero05}. A complete description of the parametric
boomerons will be the subject of a more extended report.

We performed further investigations of TWRIS stability by carrying
out a multi-scale asymptotic analysis (MAA)
\cite{pelinovsky95,buryak96}. The MAA aims to find the rate of
growth $\lambda$ (with $\lambda<<1$) of small perturbations, by
linearizing Eqs.(\ref{3wri}) around the soliton solution. This
procedure leads to a linear eigenvalue problem, whose solution can
be expressed as an asymptotic series in $\lambda$
\cite{mihalache97}. In this way, one obtains the following
condition which defines the borderline between stable and unstable
TWRIS
\begin{equation} \label{G}
G= \frac{\partial(I_{12},\overline I_{23}, \overline J
)}{\partial(\phi_1,\phi_2,V )}=0,
\end{equation}
where $G$ is the Jacobian of the constants of motion $I_{12},\overline I_{23},\overline J$
with respect to $\phi_1,\phi_2,V$. Note that in (\ref{G})
 $\overline I_{23}$ and $\overline J$ are obtained by re--normalizing
the divergent integrals (\ref{E23}) and (\ref{J}) according to the prescription
\begin{equation}
 \overline I_{23}=\frac{1}{2} \int_{-\infty}^{+\infty} (|E_2|^2 + |E_3|^2-|E_{30}|^2 )dz,
\end{equation}
\begin{equation}
\begin{split}
 \overline J=J-\frac{1}{4i}\int_{-\infty}^{+\infty} \big[(&E_3^*E_{3z}-E_3E_{3z}^* )
\bigg[\frac{|E_{30}|^2}{|E_3|^2}\bigg ] \big]dz,
\end{split}
\end{equation}
where $|E_{30}|=\lim_{z\rightarrow \infty}|E_3|$ is the asymptotic
amplitude of the kink \cite{kivshar95bis}.  Note that the
availability of exact soliton solutions allows for the analytical
calculation of the above integrals, hence of the condition
(\ref{G}), which is an extension of the well-known
Vakhitov-Kolokov criterion. Thus (\ref{G}) provides a sufficient
stability condition, which can only be applied under specific
constraints such as $\phi_1=\phi_2$. Indeed, in this case we find
that the condition $G=0$ leads again to the previously found
marginal stability condition $V=V_{cr}=2V_1V_2/(V_1+V_2)$.
\begin{figure}
\begin{center}
     \includegraphics[width=3.5cm]{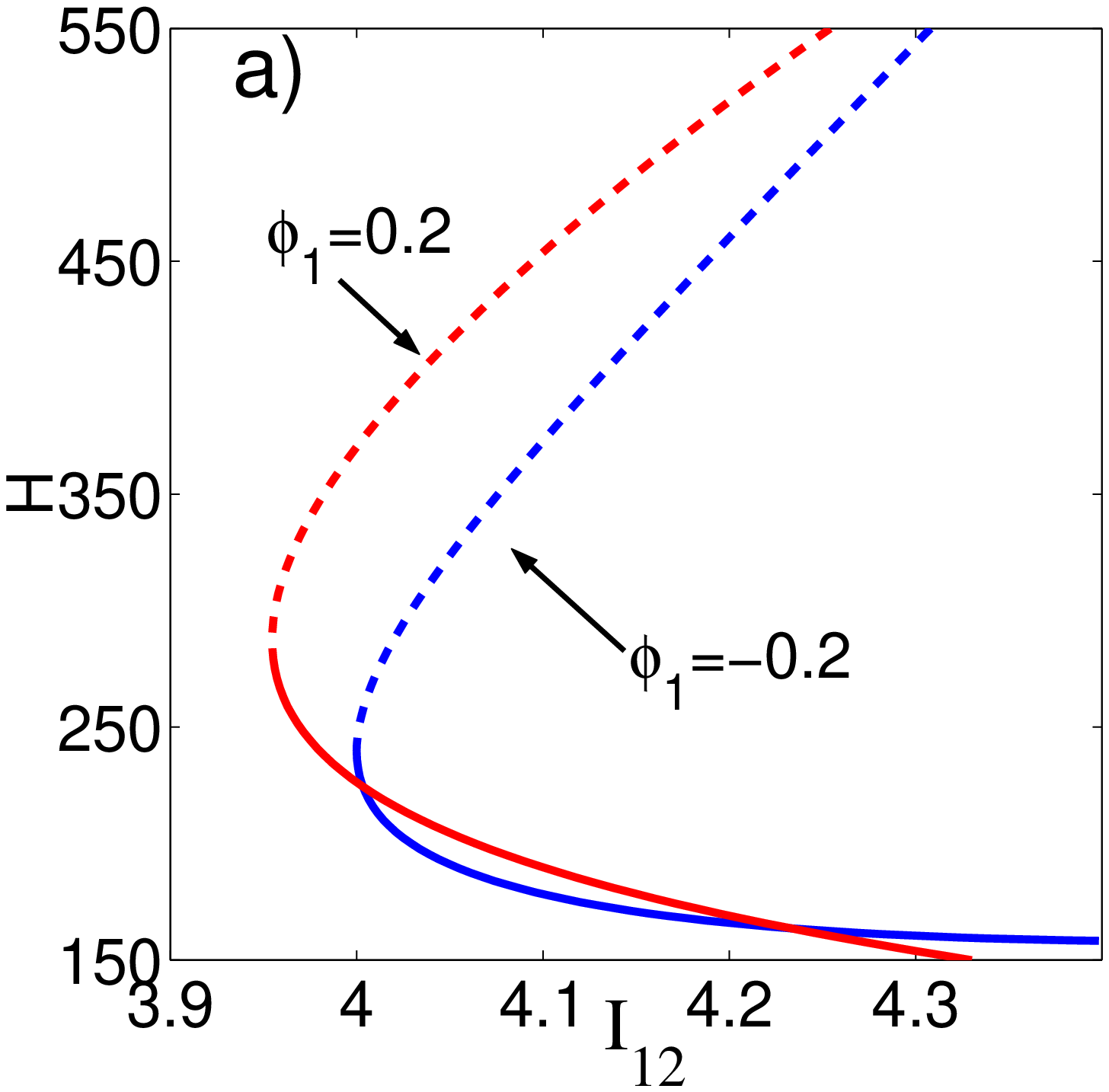}\,\,\,\,\,\,\,
     \includegraphics[width=3.5cm]{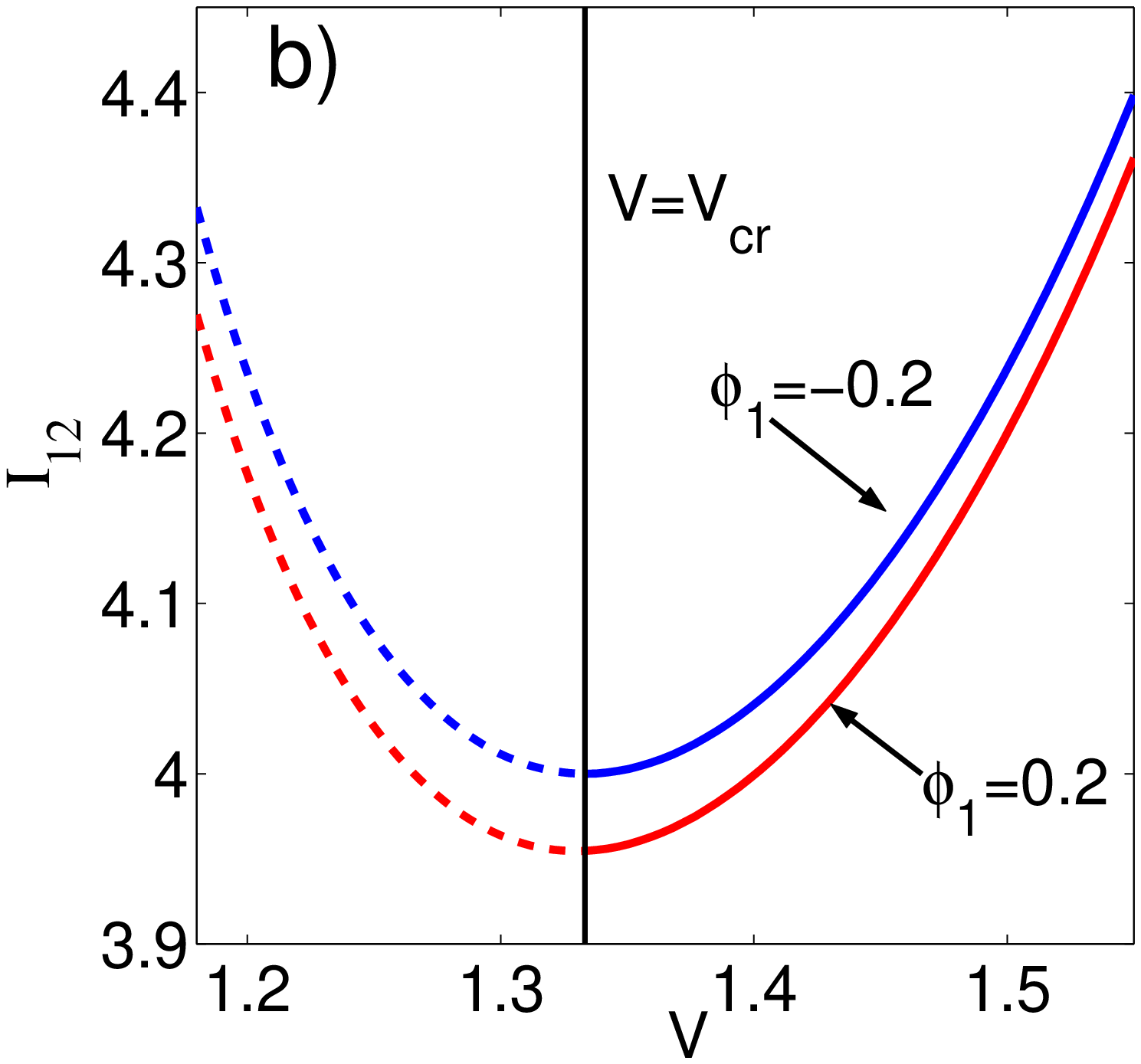}
 \end{center}
    \caption{(color online). (a) Hamiltonian $H$ versus energy $I_{12}$ and (b) $I_{12}$ versus velocity
    $V$. Characteristic velocities are  $V_1=2,V_2=1$ and $\phi_2=-0.2$.
     Dashed (solid) curves, unstable (stable) solitons.
            }\label{he}
\end{figure}

A direct insight into the global stability properties of TWRIS for
all possible values of their parameters can be obtained by means of  a
geometrical approach \cite{torner98}. Indeed, TWRIS may be
obtained as solutions of the variational problem
\begin{equation}\label{var}
\delta( H + \phi_1 I_{12} - (\phi_1 + \phi_2) \overline I_{23} -V
\overline J)=0,
\end{equation}
where $\delta$ is the Frech\'et derivative. In other words, TWRIS
represent the extrema of the Hamiltonian (\ref{H}), for a fixed value
of the energies and momentum ($V,\phi_1,\phi_2$ represent Lagrange
multipliers). Stable triplets are obtained whenever such extrema
coincide with a global minimum of $H$. Clearly, if multiple
solutions exist with the same $I_{12}, \overline I_{23}, \overline
J$, the stable solution is obtained on the lower branch of $H$. In
this framework the condition (\ref{G}) corresponds to solitons
such that the normal vector to the three--dimensional surface
$H=H(I_{12},\overline I_{23},\overline J)$ lies in the space
$H=const.$. The above geometrical considerations permit the
visualization of the stability boundaries when considering a
projection of the hyper-surface $H=H(I_{12},\overline
I_{23},\overline J)$ on the plane $(I_{12},H)$.  For example, Fig.
\ref{he}  displays the dependence of $H$ upon $I_{12}$ for the
case $\phi_1=-\phi_2$, where the criterion (\ref{G}) cannot be
applied, and in the case $\phi_1=\phi_2$. Here it is evident that
the two branches of the Hamiltonian merge exactly at $V=V_{cr}$:
at this point, the normal to the $H$ curve is also orthogonal to
the vertical axis. Interestingly enough, Fig. \ref{he} shows that
the borderline TWRIS corresponds to a minimum of the bright pulses
energy $I_{12}$ with respect to $V$. To summarize, we have shown
that different numerical and analytical methods concur in
predicting that the TWRIS stability is determined by the condition
$V>V_{cr}$.

Let us briefly discuss the experimental conditions for the observation of TWRIS in nonlinear optics. For instance, when considering a
three-wave \emph{oeo} interaction in a 5 cm-long bulk PPLN sample
with $18 \mu m$ periodicity, the field envelope carriers
$\lambda_{E1}=1.5\mu m$, $\lambda_{E2}=0.8\mu m$,
$\lambda_{E3}=1.8 \mu m$, pulse durations of about $1ps$, TWRIS
can be observed with field intensities of a few
$MW/cm^2$.

In conclusion, we have described a novel three-parameter family of
(1+1) bright-bright-dark soliton waves  as exact solutions of the
three-wave resonant interaction equation. These TWRI solitons
exhibit nonlinear phase-fronts curvatures, and exist for a given
range of their locked velocity and energy flows. Their propagation
stability has been investigated with the upshot that stable
triplets occur whenever their velocity $V$ is greater than a
certain critical value $V_{cr}$. On the other hand, unstable
solitons dynamically reshape into stable solitons with higher
velocity. The remarkable properties of these parametric solitons
may open the way to new possibilities for the control of coherent
energy transport in various physical settings.

\

\end{document}